\documentclass[conference,letterpaper]{IEEEtran}

\usepackage{amsthm}
\usepackage{comment}

\theoremstyle{definition}
\newtheorem{theorem}{Theorem}[]

\newtheorem{observation}{Observation}[]
\newtheorem{claim}{Claim}[]

\usepackage{thm-restate}

\newtheorem{corollary}{Corollary} 

\usepackage[utf8]{inputenc} 
\usepackage[T1]{fontenc}
\usepackage{url}
\usepackage{ifthen}
\usepackage{cite}
\usepackage[cmex10]{amsmath} 

\usepackage[utf8]{inputenc}
\usepackage[T1]{fontenc}
\usepackage{url}
\usepackage{ifthen}
\usepackage{cite}
\usepackage[cmex10]{amsmath}

\usepackage{amsthm}
\usepackage{algorithmic, algorithm}
\usepackage[english]{babel}
\usepackage[table]{xcolor}
\usepackage{enumitem}
\usepackage{amssymb}
\usepackage{xcolor}
\usepackage{graphicx}
\usepackage{tikz}
\usepackage[hidelinks]{hyperref}
\usepackage[font=normalsize]{caption}

\theoremstyle{definition}
\newtheorem{definition}{Definition}
\newtheorem{problem}{Problem}
\newtheorem{construction}{Construction}
\newtheorem{example}{Example}

\newcommand{\fix}[1]{{ {{#1}}}}

\newcommand{\bfc}{{\boldsymbol c}}

\newcommand{\bfs}{{\boldsymbol s}}

\newcommand{\bfu}{{\boldsymbol u}}
\newcommand{\bfv}{{\boldsymbol v}}

\newcommand{\bfgam}{\boldsymbol{\gamma}}

\newcommand{\sfA}{\mathsf{A}}
\newcommand{\sfC}{\mathsf{C}}
\newcommand{\sfG}{\mathsf{G}}
\newcommand{\sfT}{\mathsf{T}}

\newcommand{\sfM}{\mathsf{M}}

\def\CC#1#2{\ensuremath{\left(\kern-.3em\left(\genfrac{}{}{0pt}{}{#1}{#2}\right)\kern-.3em\right)}}

\begin{document}
\title{Rank Modulated Composite Encoding for Data Storage in DNA} 


\author{%
   \IEEEauthorblockN{\textbf{Tomer Cohen}\IEEEauthorrefmark{1}, \textbf{Zhiying Wang}\IEEEauthorrefmark{3}, \textbf{Eitan~Yaakobi}\IEEEauthorrefmark{1}, and \textbf{Zohar Yakhini}\IEEEauthorrefmark{1}\IEEEauthorrefmark{2}}
    \IEEEauthorblockA{\IEEEauthorrefmark{1}%
    Faculty of Computer Science, 
    Technion---Israel Institute of Technology, Israel}

    \IEEEauthorblockA{\IEEEauthorrefmark{2}%
    School of Computer Science Science, 
    Reichman University, Herzliya, Israel}

    \IEEEauthorblockA{\IEEEauthorrefmark{3}%
    Center for Pervasive Communications and Computing, 
    University of California, Irvine, CA, USA}

    \texttt{\{tomer.cohen, yaakobi, zohary\}@cs.technion.ac.il, zhiying@uci.edu\vspace{-3ex}}
    
 \vspace{-1ex}}


\maketitle


\begin{abstract}
This paper studies two problems that are motivated by combining two novel approaches, namely \emph{DNA composite} and \emph{rank modulation}. The recent approach of composite DNA takes advantage of the DNA synthesis property which generates a huge number of copies for every synthesized strand. Under this paradigm, every composite symbols does not store a single nucleotide but a mixture of the four DNA nucleotides. Instead of considering all the possible composite symbols we are interested only in the \emph{rank} of the motifs in the symbol. The first problem in this paper addresses the capacity of a channel that uses such symbols, while in the second, bounds and construction of such codes are studied.  \footnote{Funded by the European Union (DiDAX, 101115134). Views and opinions expressed are however those of the author(s) only and do not necessarily reflect those of the European Union or the European Research Council Executive Agency. Neither the European Union nor the granting authority can be held responsible for them. 
} 
\end{abstract}
\vspace{-2ex}
\section{Introduction}\label{sec:intro}

The primary challenge in making DNA data storage cost-effective is the high synthesis cost. A straightforward way to reduce this cost is by increasing the data density, measured in bits per symbol or per synthesis cycle. Standard encoding over $\sfA, \sfC, \sfG, \sfT$ allows up to $\log_2 4 = 2$ bits per symbol, but error-correction constraints often lower this limit—for example, to $\log_2 3 \approx 1.58$ bits/symbol when restricting consecutive symbols from repeating~\cite{Getal13, osti_1619517}. Introducing additional encoding symbols can improve capacity and further reduce costs.
 
\emph{Composite DNA symbols}, introduced in~\cite{anavy_DataStorageDNA_2019,augmented_encoding}, exploit inherent redundancies in DNA synthesis and sequencing processes. Unlike standard symbols that represent a single nucleotide, a composite symbol encodes a controlled mixture of the four bases and is defined by a probability vector ${p_\sfA, p_\sfC, p_\sfG, p_\sfT}$, where each $p_a$ denotes the relative abundance of nucleotide $a$, subject to $\sum p_a = 1$. For example, ${\sfM = (0.25, 0.25, 0.25, 0.25)}$ represents an equal mixture of all four nucleotides. In a sequence such as $\sfC\sfM\sfG$, the resulting oligo pool includes all $4$ possible combinations at the second position, like $\sfC\sfA\sfG$, $\sfC\sfC\sfG$, $\sfC\sfG\sfG$, and $\sfC\sfT\sfG$, with equal frequencies reflecting the underlying probabilities. During sequencing, sampling a subset of molecules from this pool enables estimation of the original base mixture.

An extension of the composite symbol model, referred to as combinatorial composite DNA, was proposed by Preuss et al.~\cite{PGYA24}. Their approach generalizes the concept of composite symbols from the nucleotide level to the shortmer level, where each symbol represents a mixture of short DNA sequences rather than individual bases. In~\cite{ZC22}, Zhang et al. investigated error-correcting codes adapted to the composite DNA model. Furthermore, several related models and their corresponding coding schemes have been explored in recent studies, including~\cite{OM1,OM2}.

It should be mentioned that the combinatorial composite DNA can be seen as a composite symbol with union distribution or in other words a \emph{subset} of composite symbols. 

The concept of \emph{rank modulation} has been excessively studied for flash memory such as in~\cite{RAMOSCH}. Codes over permutations were shown in~\cite{CORRSCH,BAMA}.

In this paper we wish to combine the theories about composite DNA and rank modulation. A composite DNA is defined by its probability distribution but a rank modulation composite symbol is defined by the rank between its motifs. For example the symbols $(0.1,0.2,0.3,0.4)$ and $(0.05,0.25,0.3,0.4)$ are different composite symbols but are the same rank modulation composite symbol since the rank between the motifs is the same. Instead of considering all the probability distributions we are interested only in the \emph{rank} of the symbols. 

We define the channel that uses rank modulated composite symbols and find codes over this alphabet.
This work studies two important aspects of the ranked composite model.
The first problem is concerned with the capacity of the channel the receives a rank modulated composite symbol as an input and in every transmission sends one motif with respect to the symbol's probability distribution. A related problem to this problem with the binomial channel was studied in~\cite{WESEL}. Another work~\cite{KYW23} studied a similar problem over regular composite symbols. In~\cite{TC24} the set of composite symbols was selected with respect to the decoding probability. Lastly,~\cite{ROM} studied the capacity of the channel with combinatorial composite symbols.

In the second problem, we first assume that the only errors in the symbol can be the rank between the motifs. Then, we bound the size of codes than can detect or correct rank errors and show a construction for such codes.

The rest of the paper is organized as follows. Section~\ref{sec:def}, introduces the rank modulated composite channel. Then we present the channel that is used for sequences over rank modulated symbols and the type of errors in this channel. Section~\ref{sec:tenscode} presents the \textbf{second} problem in the paper which asks for the bounds and constructions for codes than can detect or correct rank errors. In this section we show in Section~\ref{subsec:code_constr} the construction of the code, in Section~\ref{subsec:code_prop} its properties and its size. In Section~\ref{subsec:full_perm_constr} we show some unique codes that use rank symbols the use all the set of motifs and then generalise it for the case where only a subset of motifs is used in every symbol. Finally, we study the \textbf{first} problem, in Section~\ref{sec:cap} and calculate the channel capacity 
for several cases.

\newcommand{\motifsize}{q}
\newcommand{\wordsize}{m}

\newcommand{\setmotifs}{\mathcal{W}}
\newcommand{\rankset}[2]{\mathcal{S}_{#1,#2}}
\newcommand{\DNA}{\{A,C,G,T\}}

\newcommand{\allrankset}{{\rankset{\wordsize}{\motifsize}}}

\newcommand{\allranksetvec}{\mathcal{S}^{n}_{\wordsize,\motifsize}}

\newcommand{\inputdistname}[1]{\bfgam^{#1}}

\newcommand{\inputdistvalues}{\bfgam^{\wordsize}=(\gamma_1,\gamma_2,\dots,\gamma_{\wordsize})}
\newcommand{\distsymbol}{\gamma}

\newcommand{\inputdistsum}{\sum_{i=1}^{\wordsize}{c_i}=1}

\newcommand{\depthsize}{R}

\newcommand{\distau}{d_{\tau}}

\newcommand{\codesizedet}{\mathsf{A}_{\textbf{det}}}
\newcommand{\codesizefix}{\mathsf{A}_{\textbf{cor}}}

\newcommand{\inerror}{t}
\newcommand{\outerror}{e}

\newcommand{\rankchannel}[4]{\textbf{RMCC}(#1,#2,#3,#4)}

\newcommand{\rankchannelconsts}{\rankchannel{q}{m}{\bfgam}{R}}

\newcommand{\tauerror}{Kendall's $\tau$ error }
\newcommand{\tauerrors}{Kendall's $\tau$ errors }

\newcommand{\taupermerror}{Kendall's $\tau$ permutation error }
\newcommand{\taupermerrors}{Kendall's $\tau$ permutation errors }

\newcommand{\taudist}{Kendall's $\tau$ distance }
\newcommand{\perm}[1][\wordsize]{\mathcal{S}_{#1}}

\section{Definitions and Problem Statements}\label{sec:def}

Let $q\in \mathbb{N}$ we denote by $[q]=\{0,1,\dots,q-1\}$ the set of integers between $0$ and $q-1$.It is important to note that since this set represents a set of motifs, one can think of this set as a general set of size $q$. Given the alphabet size $q$, for example $\motifsize=4$ for DNA, the per-symbol capacity is limited by the alphabet size. A recent approach to break this limitation, first studied in~\cite{PGYA24}, used the so-called \emph{combinatorial composite symbols}. The set of combinatorial composite symbols of size $\wordsize$ is the set of all subsets of $[q]$ of size $\wordsize$, i.e., 
$
{\Sigma_{\textbf{comb}}=\{A\subseteq[q] : |A|=\wordsize\}}.
$
It holds that
$
{|\Sigma_{\textbf{comb}}|=\binom{q}{\wordsize}}.
$
A more general idea of this approach was first studied in~\cite{anavy_DataStorageDNA_2019,augmented_encoding},
with the name of
\emph{composite symbols}. Unlike combinatorial composite symbols that are defined using \emph{sets}, the composite symbols set is defined as all the probability distributions over the set $[q]$, i.e., 
$
{\Sigma_{\textbf{comp}}=\{\bfgam\in \Delta_{q}\}},
$
where $\Delta_{q}$ is the $(q-1)$-dimensional probability simplex, i.e.,
${
\Delta_{q}=\left\{(p_1,p_2,\dots,p_q)\in \mathbb{N}^q\text{ : } \sum_{i=1}^{q}{p_i}=1\right\}.
}$
The use of composite symbols was first introduced over DNA nucleotides $\DNA$. One can note that there are infinitely many composite symbols. According to this definition one can view a combinatorial composite symbol as a composite symbol with uniform distribution over a subset of the original alphabet.  
 

In previous works~\cite{TC24,KYW23}, the main goal was to study how to choose a subset of composite symbols to minimise the decoding failure error or to maximise the capacity of the channel that uses composite symbols. This approach coincides with the real use of composite symbols where only a limited number of symbols are used and their distribution is often limited to a finite set of probabilities. Finding error-correcting codes with composite symbols was studied in~\cite{OM1, OM2}.





In this paper we introduce a new approach that combines the benefits of these two approaches. Namely, we wish to benefit from the simplicity of combinatorial composite while increasing the number of composite symbols. Instead of considering all the probability distributions we are interested only in the \emph{ranking} between the symbols of every composite symbol. 
\begin{example}\label{ex:rank}
    For $q=4$ we can consider the alphabet to be the DNA nucleotides $\DNA$. When using composite symbols the set of composite symbols consists of all probability distributions over $\DNA$, i.e., $\Sigma_{\textbf{comp}}=\{(p_A,p_C,p_G,p_T):p_b\geq0,b \in \{A,C,G,T\},p_A+p_C+p_G+p_T=1\}$. 
    When using combinatorial composite symbols of size $2$ we get all the subsets of size $2$ over $\DNA$, i.e.,
    $
    \Sigma_{\textbf{comb}}=\{AC,AG,AT,CG,CT,GT\}.
    $
    A symbol in $\Sigma_{\textbf{comb}}$, for example, $AC$, can be realized by the uniform probability symbol in $\Sigma_{\textbf{comb}}$ for the symbols $\{A,C\}$.

    
    
\end{example}

Combining the ideas of composite symbols and combinatorial composite symbols, we consider the following alphabet as the channel input. Fix two positive integers $m \le q$.
We denote by 
$
\rankset{\wordsize}{\motifsize}
$
the \emph{partial permutations},
which contains all ordered subsets of $[q]$ of size $\wordsize$.
It holds that
$
|\rankset{\wordsize}{\motifsize}|=\frac{{\motifsize}!}{({\motifsize}-{\wordsize})!}.
$
\fix{For $m=q$ we denote by $\perm$ the set of all permutations of size $m$}

In the context of DNA storage, the set $[q]$ is called the  \emph{motifs}. The value $q$ denotes  the \emph{number of motifs}.  
The \emph{size of the ranked symbol} is the number of the partial permutations.

\begin{example}
    For $\motifsize=4$ we can view $[q]$ as the DNA symbols $\DNA$. For $\wordsize=2$ we have the set of size $\frac{4!}{2!}=12$ that contains $\rankset{2}{4}=\{AC,CA,AG,GA,AT,TA,CG,GC,
        CT,TC,GT,TG\}$.
        
\end{example}

Let $\pi\in \rankset{\wordsize}{\motifsize}$ we denote by $\xi(\pi)$ a vector than contains the motifs in $\pi$ in increasing order of their motif value. Furthermore, we define $\pi\downarrow$ to be the permutation $\pi$ after renaming the motifs such that $\xi(\pi)_i$ is renamed to $i$. One can observe that 
$
\pi\downarrow\in \perm$.

\begin{example}
    For $\motifsize=5$ and $\wordsize=3$ it holds that $135,251\in \rankset{\wordsize}{\motifsize}$. Furthermore, $\xi(135)=(1,3,5)$ and $\xi(251)=(1,2,5).$
    Finally, $135\downarrow=123$ and $251\downarrow=231$. 
\end{example}

As we have shown in Example~\ref{ex:rank} for a ranked composite symbol we need to choose a probability distribution that generates the ranking of the symbols. In this work we assume that when using symbols of length $m$ all the probability distributions are the same. For $\wordsize\in \mathbb{N}$ we define the \emph{symbol inner distribution} to be a fixed probability distribution denoted by ${\inputdistvalues}$. One can note that for $\wordsize=2$ it holds that $\inputdistname{2}=(\distsymbol_1,\distsymbol_2)$ so we can refer to $\inputdistname{2}$ as $\distsymbol_1$.

In Example 2, $\gamma^2 = (\gamma_1,1-\gamma_1)$ where $\gamma_1\in [0,\frac{1}{2})$ can represent the symbol $AC$ since the probability of $A$ is less than $B$'s. 
Next the \emph{rank modulated composite channel} is formally introduced. In~\cite{ROM}, the authors studied the same channel only over combinatorial composite symbols. Our channel uses symbols of ranked motifs instead of sets of motifs.
\begin{definition}\label{def:channel}
    Let $\motifsize,\wordsize\in \mathbb{N}$ be the numebr of motifs and the size of the ranked symbols. 
    Let ${\inputdistvalues}$ be a probability distribution such that
    $
    0\leq \gamma_1<\gamma_2<\dots<\gamma_m\leq1.
    $
    Let $R$ be the number of transmissions on the channel, it will be referred as the \emph{coverage depth} of the channel similarly to the coverage depth of DNA synthesis that refers to the number of times a nucleotide is read during sequencing. 
    We denote the channel by $\rankchannelconsts$.

    \textbf{Input:}
    A partial permutation $\pi=(\pi_1,\pi_2,\dots,\pi_m)\in\allrankset $ 

    \textbf{Transmission:}
    The channel transmits $R$ symbols with repetitions from the $m$ symbols associated with the input $\pi$. The symbol in each transmission is selected i.i.d with respect to the probability distribution $\inputdistvalues$.

    \textbf{Output:}
    The output alphabet of the channel is denoted by $\mathcal{Y}$ and  is equal to all the multi-sets of size $R$ over $[q]$. It will be easier in this paper to view $\mathcal{Y}$ as the set of all vectors of length $q$ such that the value in index $i$ represents the number of occurrences of the motif $i$, i.e.,
    $
    \Psi_R=\left\{\bfv\in\mathbb{N}^{q} \ : \ \sum_{i=0}^{q-1}{v_i}=R\right\}.
    $
    Specifically, when $v_i=0$ for all $i$ not in the set associated with $\pi$, and the channel transition probability is: ${P((v_0,\dots,v_{q-1})|(\pi_1,,\dots,\pi_m))= \binom{R}{v_{\pi_1}, \dots, v_{\pi_m}} \prod_{i=1}^m \gamma_i^{v_i}}$.
    \end{definition}
    Denote by $\mathsf{cap}(q,m,R, \bfgam)$ the capacity of the channel $\rankchannelconsts$. Denote by $\mathsf{cap}(q,m,R)$ the maximal capacity that can be achieved in the channel $\rankchannelconsts$ for all possible choices of the symbol inner distribution $\gamma$, i.e.,
$
\mathsf{cap}(q,m,R)=\max_{\bfgam}{\mathsf{cap}(q,m,R, \bfgam)}.
$
Lastly, let $\Gamma(q,m,R)$ be the set 
of probability distributions  ${\bfgam=(\gamma_1,\gamma_2,\dots,\gamma_m)}$ that maximises the channel's capacity, i.e.,
$
\Gamma(q,m,R)=\arg\max_{\bfgam}{\mathsf{cap}(q,m,R, \bfgam)}.
$
The first problem we seek to explore under this study is described as follows. 
\begin{problem}\label{pb:cap}
    For every $\motifsize,\wordsize,\depthsize$ compute the value of $\mathsf{cap}(q,m,R)$ and find the distribution $\Gamma(q,m,R)$.
\end{problem}






For the next problem we wish to study error-correcting codes over the rank modulated composite channel. For that, we are interested in \emph{sequences} over partial permutations. The length of the sequences is denoted by $n$. Following Definition \ref{def:channel}, we can assume that the output of the channel is also a sequence of partial permutations. Accordingly, one could consider the following errors,
\begin{enumerate}
    \item Change in order, example: sending $352$ and receiving $235$.
    \item Deletions, example: sending $352$ and receiving $52$.
    \item Substitutions, for example sending $352$ and receiving $452$.
\end{enumerate}

In this paper, we study sequences over partial permutations of the same constant size $\wordsize$. Therefore, we can read until we have all the required motifs since $\wordsize$ is known to the reader. We assume that motifs are sent with respect to $\bfgam$ so there are no substitutions and we ignore errors during sequencing which change the motif value. Hence, we are left with the first type of errors that modify the ranking of the motifs, which is the main focus of this part of the paper.
It is possible to note that this family of errors correspond to the Kendall's $\tau$-metric, which is denoted by $\distau(\cdot,\cdot)$.

\begin{example}\label{ex:tau_dist}
    For $\wordsize=3$ it holds that
    $123,231\in \perm$. Two inversions are needed to transform from $123$ to $231$ and therefore,
    $
    \distau(123, 231)=2.
    $
    For $\motifsize=5$ from Ex. we have that $135,251,351\in \rankset{\wordsize}{\motifsize}$. Since $\xi(135)\neq \xi(251)$ it holds that
    $
    \distau(135, 251)=\infty.
    $
    Since $\xi(135)= \xi(351)$ it holds that
    $
    {\distau(135, 351)=\distau(135\downarrow, 351\downarrow)=\distau(123, 231)=2}.
    $
\end{example}
It is said that $u\in {\rankset{\wordsize}{\motifsize}} $ experienced $t$ \tauerrors from $v\in{\rankset{\wordsize}{\motifsize}}$ if 
$\distau(u,v)=t.$
Let $\bfu,\bfv\in{\rankset{\wordsize}{\motifsize}}^{n}$ be two vectors of partial permutations. It is said that $\bfu$ experienced $(t,e)$-Kendall's $\tau$ permutation errors, if for at most $e$ indices, the partial permutation $u_i$ experienced at most $t$ \tauerrors from $v_i$.

\begin{example}
    For $n=2,\motifsize=5, \wordsize=3$ it holds that $(134,135),(134,351)$ are vectors in $\allranksetvec$. Following Example~\ref{ex:tau_dist} it holds that $\distau(135, 351)=2$ and $\distau(134, 134)=0$.
    Therefore, the vector $(134,351)$ experienced $(1,2)$-\taupermerrors from the vector $(134,135)$. 
\end{example}

We say that a code over $\allranksetvec$ is a \emph{$(t,e)$-error detecting code} if it can detect any $(t,e)$-Kendall's $\tau$ permutation errors. 
Similarly, a code over $\allranksetvec$ is a \emph{$(t,e)$-error correcting code} if it can correct any $(t,e)$-Kendall's $\tau$ permutation errors. 
Lastly, we denote by $\codesizedet(n,\motifsize,\wordsize,\inerror,\outerror), \codesizefix(n,\motifsize,\wordsize,\inerror,\outerror)$, the largest length-$n$ $(t,e)$-error detecting, correcting code, respectively. Codes that correct these types of errors will be referred as \emph{Kendall Permutation Codes}.

\begin{problem}\label{pb:codes}
    For every $n,\motifsize,\wordsize,\inerror,\outerror\in\mathbb{N}$ compute the size of $\codesizedet(n,\motifsize,\wordsize,\inerror,\outerror)$ and $\codesizefix(n,\motifsize,\wordsize,\inerror,\outerror)$ and find efficient constructions for such codes.
\end{problem}

First we study Problem~\ref{pb:codes}. In the following section we present a general construction for such codes.

\newcommand{\dmin}{d_\textbf{min}}
\newcommand{\balltau}[1]{B^{\tau}_{#1}}
\newcommand{\spheretau}[1]{\mathcal{S}^{\tau}_{#1}}

\newcommand{\partperm}{\mathcal{R}}
\newcommand{\partitionsize}{\ell}

\newcommand{\outercode}{\mathcal{C}^\textbf{outer}}

\newcommand{\tensorcodename}{ tensor permutation code }

\section{Constructions of Kendall Permutation Codes}\label{sec:tenscode}

\subsection{Constructions of Kendall Permutation Codes}\label{subsec:code_constr}

In this subsection we present a construction of Kendall Permutation Codes over partial permutations. 
The idea behind this construction is based on the work of tensor product codes using two linear codes~\cite{TENS}, however, in our case the set of permutations does not constitute a linear space and thus cannot be easily used with the tensor product framework. We will first construct codes for permutations, i.e., $\wordsize=\motifsize$, and then extend for partial permutations ($\wordsize>\motifsize$) in the final subsection.

First we define a crucial building block for our code. Let $\motifsize,\wordsize\in\mathbb{N}$. We already defined $\allrankset$ to be all the partial permutations over $[q]$ of size $\wordsize$. Let $A\subseteq \allrankset$ be a subset of partial permutations. We denote by $\dmin(A)$ the minimum \taudist between any two partial permutations in $A$, i.e., 
${
\dmin(A)=\min_{a\neq a'\in A}{\distau(a,a')}.
}$
A family of sets ${\partperm=A_0,A_1,\dots,A_{\partitionsize-1}\subseteq\allrankset}$ is called a \emph{partial partition} of size $\partitionsize$ of $\allrankset$ if for every $i\neq j$ it holds that
$
A_i\cap A_j=\phi.
$
For $q=m$ we also say that it is a partial partition of $\perm$.\fix{From now on in this paper the notation of $\partperm$ represents a general partial partition $A_0,A_1,\dots,A_{\partitionsize-1}\subseteq\allrankset$ of size $\partitionsize$ of $\allrankset$.}
A partial partition $\partperm$
is called a \emph{partition} if 
$
\cup_{i}{A_i}=\allrankset.
$
The minimum \taudist of a partition $\partperm$
is defined by 
$
{\dmin(\partperm)=\min_{i}{\dmin(A_i)}}.
$
For every vector of partial permutations $\bfc\in{\allranksetvec}$ we denote by $\Lambda(s)$ the vector of indicators of the partial partition $\mathcal{R}$, i.e., in index $i$ we have the value $j$ if $\bfs_i\in A_j$.  
If the permutation is not in any set we denote it by ``$?$''. We omit the partial partition $\mathcal{R}$ from the notation of $\Lambda$ as it will clear from the context. 

\begin{example}\label{ex:lambda}

For $\motifsize=5,\wordsize=3$ the following partial partition $\partperm_1$ has minimum  \taudist $3$, ${A_0=\{123,321\},A_1=\{132,231\}}$.
    It holds that $\Lambda((123,123,231))=(0,0,?),\Lambda((321,123,321))=(0,0,0),\Lambda((231,132,321))=(?,1,0)$.
    
\end{example}

Now we are ready to present our construction that follows the ideas of tensor product codes~\cite{TENS}. 

\begin{construction}\label{cnstr:TPC}
Let $\partperm$
be a partition with minimum \taudist $d_{\textbf{in}}$ that will be referred as the \emph{inner code}. Let $\outercode$ be a length-$n$ code over $[\partitionsize]$ 
with minimum Hamming distance $d_{\textbf{out}}$ 
that will be\fix{referred to as} the \emph{outer code}.    
A vector of partial permutations $\bfc\in{\allranksetvec}$ is a codeword in the tensor permutation code
$\mathbf{TPC}(\partperm,\outercode)$ if and only if 
$
\Lambda(\bfs)\in\outercode.
$
\end{construction}
A code that is constructed by Construction~\ref{cnstr:TPC} will be\fix{referred to as} \emph{tensor permutation code}.


\begin{example}\label{ex:code_3_3}
    We use the same partition $\partperm_1$ from Example~\ref{ex:lambda} with minimum \taudist $3$.
    The code $\outercode_1=\{11111,00011\}$ is a binary code of length $n=5$ with minimum Hamming distance $d_{\textbf{out}}=3$. 
The set of vectors $\bfc\in{\allranksetvec}$ such that $\Lambda(\bfc)=00011$ is the set $A_0\times A_0\times A_0\times A_1\times A_1$, totaling $32$ words. Some of these vectors are $(123,123,123,132,132)$,$(123,123,123,132,231)$,
$(123,123,123,231,132)$, and $(123,123,123,231,231)$. 
    Similarly, all the vectors $\bfc\in{\allranksetvec}$ such that $\Lambda(\bfc)=11111$ is the set of $A_1\times A_1\times A_1\times A_1\times A_1$.   
    Since $|A_0|=|A_1|$ it holds that the total size of the code is $|\outercode_1| |A_1|^5 = 2\times2^5=64$.
\end{example}



\newcommand{\disham}{d_H}
\newcommand{\dcodein}{d_{\textbf{in}}}
\newcommand{\dcodeout}{d_{\textbf{out}}}

\subsection{Properties and Size of Tensor Permutation Codes}\label{subsec:code_prop}

In the following theorems we go over the properties that the tensor permutation codes inherit from their inner and outer codes.
Let $\partperm$
be an inner code with minimum \taudist $\dcodein$. Let $\outercode$ be an outer code over $[\partitionsize]$ of length $n$ 
with minimum Hamming distance $\dcodeout$.
First we begin with two simple observations.

\begin{observation}\label{obs:lambda_change}
    For every two codewords $\bfc_1,\bfc_2\in\mathbf{TPC}(\partperm,\outercode)$ exactly one of the following occurs: $\disham(\Lambda(\bfc_1),\Lambda(\bfc_2))=0$ or $\disham(\Lambda(\bfc_1),\Lambda(\bfc_2))\geq \dcodeout$.
\end{observation}

\begin{observation}\label{obs:index_change}
Let $\bfc\in\mathbf{TPC}(\partperm,\outercode)$ and $e\leq n$. Let $\bfc'\in{\allranksetvec} $ be a vector of partial permutations that experienced $(\dcodein-1,e)$-\taupermerrors from $\bfc$. It holds that in every index $i$ the partial permutation $c'_i$ experienced \tauerror from $c_i$ if and only if $\Lambda(\bfc)_i\neq\Lambda(\bfc')_i$.
\end{observation}

\begin{theorem}\label{th:prop_detect}
    the tensor permutation code $\mathbf{TPC}(\partperm,\mathcal{C}_\textbf{outer})$ is a $(\dcodein-1,\dcodeout-1)$-error detecting code.
\end{theorem}


\begin{example}
    Using the construction of the tensor permutation code in Example~\ref{ex:code_3_3} and Theorem~\ref{th:prop_detect} it holds that $\mathbf{TPC}(\partperm_1,\outercode_1)$ is a $(2,2)$-error detecting code.
    We exemplify the detection process. Let $\bfc=(321,321,321,132,132)\in\mathbf{TPC}(\partperm_1,\outercode_1)$. The vector $\bfc_1=(132,321,321,132,213)$ experienced $(2,2)$-\taupermerrors from $\bfc$. It holds that
    $
    \Lambda((312,321,123,132,213))=(1,0,0,1,?),
    $
    and hence,  $\bfc_1\notin\mathbf{TPC}(\partperm_1,\outercode_1)$.
    The vector $\bfc_2=(321,321,231,132,132)$ experienced $(1,1)$-\taupermerrors from $\bfc$. It holds that
    $
    \Lambda((321,321,231,132,132))=(0,0,1,1,1),
    $
    but $(0,0,1,1,1)\notin\outercode_1$,  hence, $\bfc_2\notin\mathbf{TPC}(\partperm_1,\outercode_1)$.
\end{example}

\begin{theorem}
 The tensor permutation code $\mathbf{TPC}(\partperm,\mathcal{C}_\textbf{outer})$ is a $(\left \lfloor{\frac{\dcodein-1}{2}}\right \rfloor,\left \lfloor{\frac{\dcodeout-1}{2}}\right \rfloor)$-error correcting code. 
\end{theorem}

\begin{example}
    Using the construction of the tensor permutation code in Example~\ref{ex:code_3_3} and Theorem~\ref{th:prop_detect} it holds that $\mathbf{TPC}(\partperm_1,\outercode_1)$ is a $(1,1)$-error correcting code.
    We exemplify the correction process. Let $\bfc=(321,321,321,132,132)\in\mathbf{TPC}(\partperm_1,\outercode_1)$. The vector $\bfc_2=(321,321,231,132,132)$ experienced $(1,1)$-\taupermerrors from $\bfc$. It holds that
    $
    \Lambda((321,321,231,132,132))=(0,0,1,1,1),
    $
    and there exists only one codeword in $\outercode_1$ with Hamming distance at most $1$ from $(0,0,1,1,1)$, which is the codeword $(0,0,0,1,1)\in\outercode_1$. Now it is known that the \tauerror occurred in index $2$. There exists only one partial permutation in $A_0$ with \taudist at most $1$ from $231$, which is the partial permutation $321$. Finally it is known that the original codeword was $(321,321,321,132,132)$ which is exactly $\bfc$.
\end{example}


\newcommand{\numof}[1]{\##1}


Next, we analyse the size of tensor permutation codes. The main difference between the analysis of different codes will be its inner code, i.e., the partial partition of the set of partial permutations.
Let $\bfv$ be a vector over $[\partitionsize]$ of length $n$. Denote by $\numof{i}(\bfv)$ the number of occurrences of $i$ in $\bfv$. The following theorem provides the size of a general tensor permutation code and a tensor permutation code over a partition such that all the sets are of the same size.

\begin{theorem}\label{thm:size}
    Let $\partperm$
    be a partial partition and let $\outercode$ be a code over $[\partitionsize]$ of length $n$. It holds that
    $
    |\mathbf{TPC}(\partperm,\outercode)|=\sum_{\bfv\in\outercode}{\prod_{i=0}^{\partitionsize-1}{|A_i|^{\numof{i}(\bfv)}}}.
    $
    If for every $i$ it holds that
    $
    |A_i|=A
    $
    then it holds that
    $
    |\mathbf{TPC}(\partperm,\outercode)|=A^n|\outercode|.
    $
\end{theorem}

\newcommand{\binpart}{\mathcal{R}_{\text{Parity}}}

\newcommand{\tauweight}{Kendall's $\tau$ weight }

\subsection{Special Constructions 
}\label{subsec:full_perm_constr}
In this subsection we will show several partitions of $\perm$. At the end we will show the relationship between tensor permutation codes over permutations and over partial permutations.  
We use the same notations as in~\cite{KNUTH} for some properties of permutations.
We denote by\fix{$\text{id}=(12\dots m)$} the identity permutation.
We say that a permutation $\pi\in \perm$ has \tauweight $k$ if it holds that
$
\distau({\text{id}},\pi)=k.
$
We denote by $I_m(k)$ the set of all permutations of \tauweight exactly $k$.
First we find a partition with minimum \taudist $2$. Let 
$\binpart=A_0,A_1$ be the partition of all the permutations according to their \tauweight parity. One can note that
$
A_0\cap A_1=\phi.
$
It is well known that the partition $\binpart=A_0,A_1$ has minimum \taudist $2$ and the sets are of equal sizes.
We denote by $A_q(n,d)$ the size of a largest code over $[q]$ of size $n$ with minimum Hamming distance $d$.
Let $\outercode_{\text{bin}}$ be a largest binary of size $n$ with minimum Hamming distance $\dcodeout$.
\begin{corollary}
    Let $\binpart$ be the inner code.
    The code $\mathbf{TPC}(\binpart,\outercode_{\text{bin}})$ is a $(1,\dcodeout-1)$-detecting code of length $n$ of size
    $
    \frac{(\wordsize!)^{n}}{2^{n}}{A_2(n,\dcodeout)}.
    $
    Furthermore, it holds that
    $
    \codesizedet(n,\wordsize,\wordsize,1,\dcodeout-1)\geq\frac{(\wordsize!)^{n}}{2^{n}}{A_2(n,\dcodeout)}.
    $
\end{corollary}

Lastly, the next theorem connects between tensor permutation codes over permutations and partial permutations.
\begin{theorem}
For every $n,\motifsize,\wordsize,\inerror,\outerror\in\mathbb{N}$ such that $\wordsize\leq\motifsize$ it holds that $\codesizedet(n,\motifsize,\wordsize,\inerror,\outerror)\geq {\binom{q}{m}}^n\codesizedet(n,\wordsize,\wordsize,\inerror,\outerror)$ and $\codesizefix(n,\motifsize,\wordsize,\inerror,\outerror)\geq {\binom{q}{m}}^n\codesizefix(n,\wordsize,\wordsize,\inerror,\outerror)$.
\end{theorem}

\vspace{-1ex}
\section{The Rank Modulated Composite Channel's Capacity}\label{sec:cap}
\vspace{-1ex}

Note that in the rank modulated composite channel $\rankchannelconsts$ the input consists of ${\frac{q!}{(q-m)!}}$ partial permutations, while the probability distribution of each partial permutation is $\bfgam$. In order to compute the channel's capacity one should also consider the probability distribution \emph{over the partial permutations}, i.e., all possible channel inputs. Since the channel is symmetric and the only difference between the partial permutations is their motifs' values, the first claim states that the capacity achieving input distribution (CAID) to the channel is the uniform distribution over all the partial permutations.\vspace{-1ex}
\begin{claim}\label{th:uniform_max}
For every $1\leq m\leq q\in\mathbb{N}$, $R\in\mathbb{N} $ and $\bfgam\in \Delta_q$ it holds that the uniform probability distribution over all the input symbols is a CAID. 
\end{claim}
Next, one can note that using the central limit theorem, when the coverage depth increases one can distinguish between the different partial permutations for every $\bfgam$ such that all the $m$ motifs have positive probabilities. Hence, the capacity approaches the logarithm of the input size.
\begin{claim}
For every $m\leq q$ it holds  $
\lim_{R\rightarrow\infty}\mathsf{cap}(q,m,R)=\log_2{\frac{q!}{(q-m)!}}.
$
\end{claim}
The following theorem calculates the capacity and the probability distribution $\bfgam$ for several special cases. 
\begin{theorem}
The following properties hold.
\begin{enumerate}
    \item For every $R\in\mathbb{N}$ it holds that $\Gamma(q=2,m=2,R)=(0,1)$ and $\mathsf{cap}(q=2,m=2,R)=\log_2(2)=1$. 
    \item For every $q\in\mathbb{N}$ it holds that $\Gamma(q,m=2,R=1)=(0,1)$ and $\mathsf{cap}(q,m=2,R=1)=\log_2(q)$. 
    \item For every $q,m\in\mathbb{N}$ it holds that $\Gamma(q,m,R=1)=e_{m,m}$ and $\mathsf{cap}(q,m,R=1)=\log_2(q)$.
    \item It holds that $\Gamma(q=3,m=2,R=2)=(0,1)$ and $\mathsf{cap}(q=3,m=2,R=2)=\log_2(3)$.
\end{enumerate}
\end{theorem}

The last theorem shows that for partial permutation of size $2$, if the alphabet size is large enough, then the best probability distribution $\bfgam$ is uniform over the two motifs. Since we normally use alphabets of size $4$ for DNA, the use of partial permutations can still be useful, for example with $q=4$, i.e. DNA and partial permutations of size $\wordsize=2$ with coverage depth $R=2$ it holds that $\Gamma(q=4,m=2,R=3)=(0.2,0.8)$. Furthermore it holds that $\mathsf{cap}\left(q,m=2,R=2,\left(0.2, 0.8\right)\right)=2.67>2.5=\mathsf{cap}\left(q,m=2,R=2,\left(\frac{1}{2},\frac{1}{2}\right)\right)$. Hence, partial permutations improved the capacity over the case of combinatorial composite symbols.
\begin{theorem}
There exists $Q\in\mathbb{N}$ such that for every $q>Q$ it holds that
$
\mathsf{cap}(q,m=2,R=2)=\mathsf{cap}\left(q,m=2,R=2,\left(\frac{1}{2},\frac{1}{2}\right)\right)
$
\end{theorem}

\section*{Acknowledgments}
The collaboration and discussions of this paper were originated in Dagstuhl Seminar 24511~\cite{DSTU}.

\end{document}